# Model-driven engineering IDE for quality assessment of data-intensive applications

Marc GIL
Prodevelop SL (Spain)
mgil@prodevelop.es

Christophe JOUBERT
Prodevelop SL (Spain)
cjoubert@prodevelop.es

Ismael TORRES
Prodevelop SL (Spain)
itorres@prodevelop.es

## ABSTRACT

This article introduces a model-driven engineering (MDE) integrated development environment (IDE) for Data-Intensive Cloud Applications (DIA) with iterative quality enhancements. As part of the H2020 DICE project (ICT-9-2014, id 644869), a DICE framework is being constructed and it is composed of a set of tools developed to support the DICE methodology. One of these tools is the DICE IDE which acts as the front-end of the DICE methodology and plays a pivotal role in integrating the other tools of the DICE framework. The DICE IDE enables designers to produce from the architectural structure of the general application along with their properties and QoS/QoD annotations up to the deployment model. Administrators, quality assurance engineers or software architects may also run and examine the output of the DICE tools in addition to the designer in order to assess the DIA quality in an iterative process.

**Categories and Subject Descriptors**

Software and its engineering → Development frameworks and environments

Software and its engineering → Software system models → Model-driven software engineering

## 1. INTRODUCTION

Recent years have seen the rapid growth of interest for cloud-based enterprise applications built on top of data-intensive technologies (DIA). However, the heterogeneity of the technologies and software development methods in use is still a challenge for researchers and practitioners.

The DICE Project [1] tries to solve the previous challenge by defining a methodology and a set of tools that will help software designer reasoning about reliability, safety and efficiency of data-intensive applications. The main goal of DICE Project is to define a Model-Driven Engineering (MDE) [2] approach and a Quality Assurance (QA) toolchain for developing DIA that leverage Big Data technologies hosted in private of public clouds.

The pivotal tool of the DICE project is the DICE IDE. It integrates all the tools of the DICE platform and it gives support to the DICE methodology. The DICE IDE is an integrated development environment tool for MDE where a designer can create models to describe data-intensive applications and their underpinning technology stack.

The aim of these tools is to put the developer in conditions to deliver an application to market in a short period of time, focusing on quality assessment, architecture enhancement, continuous testing and agile delivery.

## 2. DICE IDE DESCRIPTION

The DICE IDE aims at offering the ability to specify DIAs through UML models. From these models, the toolchain guides the developer through the different phases of quality analysis, formal verification being one of them.

The DICE IDE is based on Eclipse Neon 4.6, which is the de-facto standard for the creation of software engineering models based on the MDE approach. DICE customizes the Eclipse IDE with suitable plug-ins that integrate the execution of the different DICE tools, in order to minimize learning curves and simplify adoption. Not all tools are integrated in the same way. Several DICE integration patterns, focusing on the Eclipse plugin architecture, have been defined. They allow the implementation and incorporation of application features very quickly. Moreover, creating custom versions of DICE applications are easier and without source code modifications need. DICE Tools are accessible thought DICE Tools menu (see Figure 1)

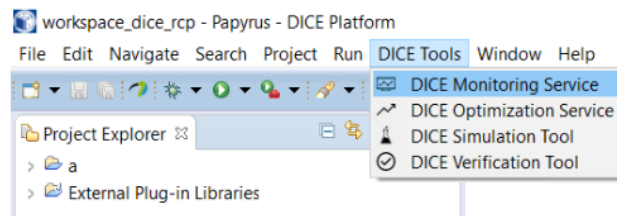

Figure 1: DICE Tools menu

### 2.1. Global Architecture

DICE focuses on delivering value to the application developer. The DICE IDE guides the developer through the DICE methodology, based on tools Cheat Sheets [3]. It initially offers the ability to specify the data-intensive application through DICE UML models stereotyped with the DICE profile[1]. From these models, the tool-chain guides the developer through the different phases of quality analysis (e.g., simulation and/or formal verification), deployment, testing, and acquisition of feedback data through monitoring data collection and successive data warehousing. Such data warehouse feds back to the IDE through the iterative quality enhancements tool-chain, detecting quality incidents and design anti-patterns based on the runtime data. Such feedback is used to guide the developer through cycles of iterative quality enhancements.

---

[1] The DICE profile provides the stereotypes and tagged values needed for the specification of data-intensive applications in UML

A conceptual architecture of the project tool-chain orchestrating the execution of the tools is shown in the Figure 2.

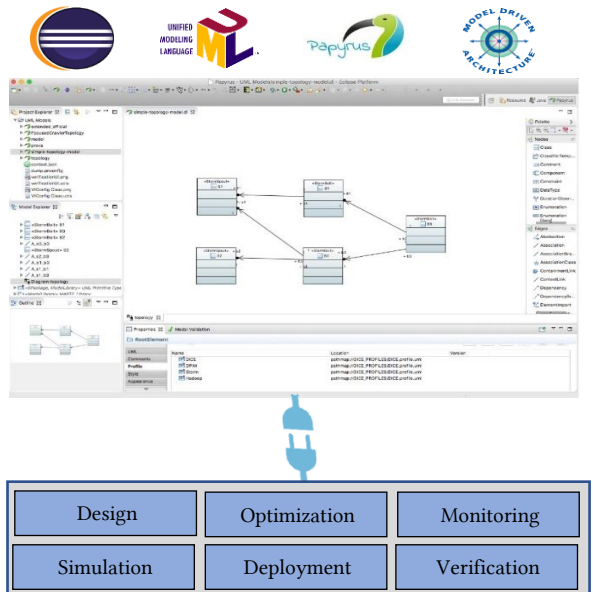

Figure 2: Global Architecture of the DICE IDE

### 2.2. MDE for Big Data Application

Models can be defined in the DICE IDE by designers in a top-down fashion, starting from platform-independent specifications of components and architecture (DICE Platform Independent Models - DPIM), through assignment of specific technologies to implement such specifications (DICE Technology Specific Models - DTSM). Finally, designers can map the application components into a concrete OASIS TOSCA-compliant (YAML format) [4] deployment specification (DICE Deployment Specific Models - DDSM), using the deployment modelling DICER tool. Model-to-model transformations can be applied to these models, to reduce the amount of manual work required from the user. To do so, the DICE IDE extends Papyrus and Papyrus Marte [5], to define the DIA UML models and apply the DICE profiles to them.

Throughout the application design, the DICE IDE offers the possibility to automatically translate certain DICE models into formal models for assessment of quality properties (efficiency, costs, safety/correctness, etc.). Each analysis requires to run dedicated tools to obtain prediction metrics. These tools are either fully integrated as Eclipse plugins in the IDE environment, or their user-interfaces are run and examined through the IDE. The simulation, optimization and verification plugins, for instance, take care of translating designer models into formal models.

Currently, the DICE IDE is in a beta version and the first complete release is scheduled for July 2017. The current version can be downloaded from the DICE GitHub repository [6], together with a tutorial, video, and further documentation, as well as all DICE tools published so far. The DICE IDE can be used as an RCP (Rich Client Platform) or a developer can build her own DICE IDE from the source code repository.

The DICE IDE has been used in tree industrial DIA use cases so far, namely a news orchestrator application, a cloud-based tax fraud detection big-data application, and a real-time maritime vessel traffic management system. Initial results are encouraging, with a number of development and application performance metrics enhanced by the use of the DICE platform. Such enhancements are a lower impact of changes in the software (Simulation Tool), faster deployment and increased number of test deployments for quality testing, more structured and easily understandable system configuration (DICE models), monitoring of application execution and quality metrics.

### 3. CONCLUSIONS

The overall goal of the DICE IDE, is to become the main access gateway for all designers and developers willing to adopt and follow the DICE methodology for building DIA applications.

Improvements provided by the use of DICE IDE in order to develop data-intensive applications are: (1) User-friendly IDE, (2) Support for most of the phases of software development cycle, (3) Updates facilities, (4) Integration with other DICE tools, (5) IDE customizations, (6) Access to remote repositories, and (7) Advanced UML modelling.

The complete release version will include an integration of further DICE tools, like in order to achieve the DICE project goals.

ACKNOWLEDGMENTS

This work was supported by the European Union under the H2020 Research and Innovation Program (DICE, grant agreement no. 644869).